\documentclass[superscriptaddress,showpacs,showkeys]{revtex4-2}
\usepackage[T2A]{fontenc}
\usepackage{color}
\usepackage{babel}
\usepackage{units}
\usepackage{bm}
\usepackage{amsmath}
\usepackage{graphicx}
\usepackage[pdfusetitle,
 bookmarks=true,bookmarksnumbered=false,bookmarksopen=true,bookmarksopenlevel=2,
 breaklinks=true,pdfborder={0 0 0},pdfborderstyle={},backref=false,colorlinks=true]
 {hyperref}
\hypersetup{citecolor=blue,linkcolor=blue,urlcolor=blue}

\begin{document}
\title{Charge asymmetry in $e^{+}e^{-}\to B^{(*)}\bar{B}^{(*)}$ processes in the vicinity of $\Upsilon(4S)$}

\author{S. G. Salnikov}
\email{S.G.Salnikov@inp.nsk.su}
\affiliation{Budker Institute of Nuclear Physics, 630090, Novosibirsk, Russia}
\affiliation{Novosibirsk State University, 630090, Novosibirsk, Russia}

\author{A. I. Milstein}
\email{A.I.Milstein@inp.nsk.su}
\affiliation{Budker Institute of Nuclear Physics, 630090, Novosibirsk, Russia}
\affiliation{Novosibirsk State University, 630090, Novosibirsk, Russia}

\date{\today}

\begin{abstract}
The effects of isotopic invariance violation in the processes
$e^{+}e^{-}\to B\bar{B}$, $e^{+}e^{-}\to B^{*}\bar{B}$, and $e^{+}e^{-}\to B^{*}\bar{B}^{*}$ are considered in the energy range between the thresholds of $B^{+}B^{-}$ and $B_{s}^{0}\bar{B}_{s}^{0}$ production.	The analysis is based on taking into account the final-state interaction in a six-channel problem. Our approach allowed us to obtain good agreement with recent Belle-II results for the ratio of the $B^{0}\bar{B}^{0}$ and $B^{+}B^{-}$ production cross sections in $e^{+}e^{-}$ annihilation in the vicinity of $\Upsilon(4S)$.
It is shown that at higher energies the ratios of the cross sections for pairs of neutral and charged
$B^{(*)}$ mesons can differ significantly from unity. A detailed
measurement of this effect will provide evidence that the nontrivial
energy dependence of the $B^{(*)}\bar{B}^{(*)}$ production cross sections is a consequence
of  interference of the particle production amplitudes in the multichannel problem.
\end{abstract}

\maketitle

\section{Introduction}
A large number of resonances have now been discovered in the annihilation and decay of various particles with the energy dependence of  production cross sections (probabilities)
which are not described by the Breit-Wigner formulas.
Attempts to describe the shape of these resonances by the sum of Breit-Wigner amplitudes
have led some authors to the conclusion that numerous new
mesons are discovered. However, it has been noted that all of these resonances have been observed near the production thresholds of a certain number of mesons (or baryons).
Therefore, a natural idea appears that new resonances with unusual
cross section energy dependence are near-threshold resonances.
The origin of these resonances  is related to large distances
of the order of 1 fm. At such distances, the main degrees of freedom of the hadronic
state that contribute to the cross section are mesons and baryons,
not quarks and gluons.

Near-threshold resonances are explained by the fact that the relative velocities of the produced particles near the threshold are small, and the particles interact for a long time. This interaction is called a final-state interaction. Large values of typical interaction  potentials result in a strong distortion of the wave functions of the produced particles.

This approach has made it possible to explain  a lot of processes in which near-threshold resonances are produced, see, e.g., \cite{Salnikov2024} and  references therein. It is shown in Ref.~\cite{Salnikov2024} that, in the presence of multiple reaction channels with identical quantum numbers and nonzero transition amplitudes between channels, a nontrivial energy dependence of the cross sections for the production of such states may arise. This statement is confirmed for   $B\bar{B}$, $B^{*}\bar{B}$, and $B^{*}\bar{B}^{*}$ production in Ref.~\cite{Salnikov2024},
and for $D\bar{D}$, $D^{*}\bar{D}$, and $D^{*}\bar{D}^{*}$  production in Ref.~\cite{Salnikov2024Production}. In these papers, using a relatively small number of parameters, good agreement was obtained between the predictions of the energy dependence of the  cross sections and the experimental data in each channel. It is important that some
peaks observed in these cross sections do not correspond to any states
having the quark structure $b\bar{b}$ or $c\bar{c}$.

Since in the case of $D^{(*)}\bar{D}^{(*)}$ production in $e^+e^-$ annihilation a large amount of experimental data exists for  both charged and neutral pairs, the effects of isotopic invariance violation due to the Coulomb field and the mass differences between charged and neutral particles have been taken into account in Ref.~\cite{Salnikov2024Production}. Furthermore,
 isotopic invariance violation has required to take into account
 the $D^{(*)}\bar{D}^{(*)}$ interaction not only in isoscalar
channels but also in isovector ones.

As for experimental data for the $B^{(*)}\bar{B}^{(*)}$ production cross sections, until recently they existed
only for the sums of the cross sections for charged and neutral particle production,
with the exception of data obtained for the ratio $\Gamma(\Upsilon(4S)\to B^{+}B^{-})/\Gamma(\Upsilon(4S)\to B^{0}\bar{B}^{0})$ at the $\Upsilon(4S)$ peak~\cite{Aubert2004a,Choudhury2023}.
Therefore, in Ref.~\cite{Milstein2021}, the effects of isotopic invariance violation have been analyzed only near the $\Upsilon(4S)$ peak.
In that case, the problem has been solved in a single-channel approximation, so  that the
amplitudes of the transitions to the $B^{*}\bar{B}$ and $B^{*}\bar{B}^{*}$ states have not been taken into account.
In Ref.~\cite{Bondar2022}, the effects of isotopic invariance violation in the decays of $\Upsilon(5S)\to B^{*+}B^{*-}$
and $\Upsilon(5S)\to B^{*0}\bar{B}^{*0}$ have also been studied in a single-channel approximation. Note that the mass of  $\Upsilon(5S)$ is also
quite close to the $B^{*}\bar{B}^{*}$ production thresholds.
In Ref.~\cite{Salnikov2024}, the cross sections for the production of $B^{(*)}\bar{B}^{(*)}$
have been  considered in the three-channel approximation, but without taking into account the effects of isotopic invariance violation.

Quite recently, fairly accurate experimental data on the production of $B^{+}B^{-}$ and $B^{0}\bar{B}^{0}$ in a wide
energy range near $\Upsilon(4S)$ have appeared~\cite{Abumusabh2025Measurements}.
Therefore, in the present paper, we study the effects of isotopic invariance violation over a wide energy range near and
above $\Upsilon(4S)$ within the framework of a multichannel problem. As a result,
the small number of parameters used in our approach has allowed us not only to describe the results of Ref.~\cite{Abumusabh2025Measurements} with good accuracy,
but also to predict the cross sections for the production of charged and neutral
$B^{(*)}\bar{B}^{(*)}$ pairs in an energy range where experimental data
are currently lacking. It is shown that the ratio of the cross sections for pair production of charged and neutral particles can differ from unity by tens of percent.
This, in our view, is an incentive for experimental study of the effects of isotopic invariance violation in the $B\bar{B}$, $B^{*}\bar{B}$, and $B^{*}\bar{B}^{*}$ channels above $\Upsilon(4S)$. Such information is important for understanding the structure of interaction of $B^{(*)}$ mesons.

\section{Description of model}

In our work, we follow the approach described in detail in Ref.~\cite{Salnikov2024Production}
in analyzing the production of $D^{(*)}\bar{D}^{(*)}$ in $e^{+}e^{-}$ annihilation.
Due to isotopic invariance violation, the wave function $\boldsymbol{\Psi}$
contains six components: $\Psi_{1}=B^{+}B^{-}$, $\Psi_{2}=B^{0}\bar{B}^{0}$,
$\Psi_{3}=(B^{+}B^{*-}+B^{-}B^{*+})/\sqrt{2}$, $\Psi_{4}=(B^{0}\bar{B}^{*0}+\bar{B}^{0}B^{*0})/\sqrt{2}$,
$\Psi_{5}=B^{*+}B^{*-}$ and $\Psi_{6}=B^{*0}\bar{B}^{*0}$. Since
$B^{(*)}\bar{B}^{(*)}$ pairs are produced in single-photon $e^{+}e^{-}$ annihilation, each component has negative $C$~parity,
orbital angular momentum $L=1$, and total angular momentum $J=1$. The total
spin of the $B^{*}\bar{B}^{*}$ pairs can be $S=0,\,2$. Due to the
lack of experimental data for individual spin states
in the $B^{*}\bar{B}^{*}$ channels, we  consider the total cross sections
for the production of these states with different spins. The production threshold
for the lightest $B^{+}B^{-}$ state is $\unit[10558{.}8]{\text{MeV}}$,
and all energies are calculated from this value. Then
the production thresholds for all states, $\Delta_{i}$, are: $\Delta_{1}=0$,
$\Delta_{2}=\unit[1]{\text{MeV}}$, $\Delta_{3}=\unit[45.3]{\text{MeV}}$,
$\Delta_{4}=\unit[46.5]{\text{MeV}}$, $\Delta_{5}=\unit[90.6]{\text{MeV}}$,
and $\Delta_{6}=\unit[91.9]{\text{MeV}}$ (see Refs.~\cite{Navas2024Review,Abumusabh2025Measurements,Hayrapetyan2026Measurements}).

The radial Schrödinger equation, which describes our six-channel
system, is
\begin{equation}
\left(p_{r}^{2}+M\mathcal{V}+\frac{L(L+1)}{r^{2}}-\mathcal{K}^{2}\right)\bm{\Psi}(r)=0\,,\quad\left(\mathcal{K}^{2}\right)_{ij}=\delta_{ij}\,k_{i}^{2}\,,\quad\mathcal{V}=\begin{pmatrix}V_{11} & V_{12} & V_{13}\\
V_{12} & V_{22} & V_{23}\\
V_{13} & V_{23} & V_{33}
\end{pmatrix}+\mathcal{V}_{C}\,,\label{eq}
\end{equation}
where $\left(-p_{r}^{2}\right)$ is the radial part of the Laplacian, $k_{i}=\sqrt{M\left(E-\Delta_{i}\right)}$,
$M=\unit[5279{,}4]{\text{MeV}}$ is the  mass of charged $B$-meson, and $E$ is the energy counted from the threshold of $B^{+}B^{-}$ production.
The wave function
\begin{equation}
\bm{\Psi}(r)=\left(\psi_{1}(r),\dots,\psi_{6}(r)\right)^{T}
\end{equation}
consists of the radial parts $\psi_{i}(r)$ of the wave functions of the states
$\Psi_{i}$, the superscript $T$ denotes transposition. The diagonal matrix
$\mathcal{V}_{C}=\left(-\alpha/r\right)\mathrm{diag}\left(1,0,1,0,1,0\right)$
describes the Coulomb interaction between charged $B^{(*)}$ mesons,
$\alpha$ is the fine structure constant. The matrices $V_{ij}$ are
symmetric blocks of dimension $2\times2$ and have the form
\begin{equation}
V_{ij}=\begin{pmatrix}U_{ij}^{(0)}(r)+U_{ij}^{(1)}(r) & -U_{ij}^{(1)}(r)\\
-U_{ij}^{(1)}(r) & U_{ij}^{(0)}(r)+U_{ij}^{(1)}(r)
\end{pmatrix},\label{eq:pot1}
\end{equation}
where the diagonal potentials correspond to transitions without changing the  charges of particles, while the off-diagonal ones describe charge-exchange processes. With this parametrization, the potentials $U_{ij}^{(0)}(r)$ correspond to the interaction of mesons in isoscalar channels, while the interaction potentials of mesons
in isovector channels are $U_{ij}^{(0)}(r)+2U_{ij}^{(1)}(r)$.
Thus, $2U_{ij}^{(1)}(r)$ is the difference between the isovector
and isoscalar potentials. Since at short distances $B^{(*)}\bar{B}^{(*)}$ pairs are produced in the isoscalar state $b\bar{b}$, and the effects of isotopic invariance violation are small, the admixture of the isovector state
arising from the final-state interaction is also small.
Thus, the isoscalar potentials $U_{ij}^{(0)}(r)$ should be
close to the interaction potentials between $B^{(*)}$ mesons found
in Ref.~\cite{Salnikov2024}, where isotopic invariance violation
has not been taken into account. However, an account for the interaction in the isovector channels is necessary for studying the effects of isotopic invariance violation.

As is well known, to describe near-threshold resonances, it is sufficient to consider
a small number of parameters, such as scattering lengths and effective ranges of
interaction (see, e.g., Ref.~\cite{Salnikov2023}).
In this case, the fine details of the potential behavior at short distances are unimportant,
and any parametrization of the potential that reproduces
these parameters can be used. The simplest parametrization is the following
\begin{equation}
U_{ij}^{(n)}(r)=u_{ij}^{(n)}\,\theta(a_{ij}^{(n)}-r)\,.\label{eq:pot2}
\end{equation}
Here $\theta(x)$ is the Heaviside function, $u_{ij}^{(n)}$ and $a_{ij}^{(n)}$
are some constants that are found by comparing the model predictions with experimental
data.

Eq.~\eqref{eq} has six linearly independent solutions that are regular at zero,
\begin{equation}
\bm{\Psi}^{(m)}=\left(\psi_{1}^{(m)}(r),\dots,\psi_{6}^{(m)}(r)\right)^{T},\qquad m=1,\dots,6\,.
\end{equation}
The method for finding these solutions is described in Ref.~\cite{Salnikov2024Production},
and the account for the Coulomb field is performed as in Ref.~\cite{Milstein2021}.
In each solution $\bm{\Psi}^{(m)}$, the $\psi_{m}^{(m)}$ component contains a converging and a diverging wave as $r\to\infty$, while all other
components contain only diverging waves. The cross sections $\sigma^{(m)}$ for the production of the states $\Psi_{m}$ have the form (see Ref.~\cite{Salnikov2024Production})
\begin{align}
 & \sigma^{(m)}=\frac{2\pi\beta_{m}\alpha^{2}}{s}\left|\sum_{i=1}^{6}g_{i}\dot{\psi}_{i}^{(m)}(0)\right|^{2}\,,\label{sec}
\end{align}
where $\beta_{m}=k_{m}/M$, $s=\left(2M+E\right)^{2}$, $g_{i}$ are some constants determining the production of the corresponding states
at short distances, $\dot{\psi}_{i}^{(m)}(r)=\partial/\partial r\,\psi_{i}^{(m)}(r)$.
Since an isoscalar state is produced at short distances,
$g_{1}=g_{2}$, $g_{3}=g_{4}$, and $g_{5}=g_{6}$. The constants $g_{1}$,
$g_{3}$, and $g_{5}$ are found by comparing the predictions for the cross sections
$\sigma^{(m)}$ with experimental data.

\section{Comparison of predictions with experimental data}
Here we consider the energy range between the production thresholds of the $B^{+}B^{-}$ and $B_{s}^{0}\bar{B}_{s}^{0}$ states. Since the production thresholds for both charged and neutral $B\bar{B}$, $B^{*}\bar{B}$, and $B^{*}\bar{B}^{*}$ pairs lie in this energy range,
to correctly describe the production cross sections, we must consider the final-state interaction, taking into account all six channels. To
determine the optimal model parameters (the radii of the potential
wells $a_{ij}^{(n)}$, the potential depths $u_{ij}^{(n)}$, and the constants
$g_{i}$), we use the following experimental data. First,
the exclusive cross sections of the processes $e^{+}e^{-}\to B\bar{B}$, $e^{+}e^{-}\to B^{*}\bar{B}$
and $e^{+}e^{-}\to B^{*}\bar{B}^{*}$, summed over charged
and neutral channels~\cite{Mizuk2021,Adachi2024Measurement}. Second,
the sum of the cross sections of all listed processes, which in the considered
energy range coincides with the cross section of the process $e^{+}e^{-}\to b\bar{b}$~\cite{Aubert2009,Dong2020}.
Third, recent experimental data for the  ratio of the cross sections
$\sigma(e^{+}e^{-}\to B^{0}\bar{B}^{0})/\sigma(e^{+}e^{-}\to B^{+}B^{-})$
in the vicinity of the $\Upsilon(4S)$ resonance~\cite{Abumusabh2025Measurements}.
Here, we use the most accurate values of the mass differences $M_{B^{0}}-M_{B^{+}}=\unit[0.495]{MeV}$ (see Ref.~\cite{Abumusabh2025Measurements}), $M_{B^{*+}}-M_{B^{+}}=\unit[45.277]{MeV}$, and $M_{B^{*0}}-M_{B^{0}}=\unit[45.471]{MeV}$ (see Ref.~\cite{Hayrapetyan2026Measurements}),
and for the mass of $M_{B^+}$ we use the PGD data~\cite{Navas2024Review}.

\tabcolsep=5pt

\begin{table}
	\centering
	\caption{Parameters of the potentials $U_{ij}^{(0)}$.}\label{tab:params0}
	\begin{tabular}{|l|c|c|c|c|c|c|}
		\hline
		& $U_{11}^{(0)}$ & $U_{22}^{(0)}$ & $U_{33}^{(0)}$ & $U_{12}^{(0)}$ & $U_{13}^{(0)}$ & $U_{23}^{(0)}$\tabularnewline
		\hline
		$u_{ij}^{(0)}\,(\mathrm{MeV})$ & $-624$ & $-356.1$ & $-595.2$ & $21.2$ & $19.1$ & $77.3$\tabularnewline
		$a_{ij}^{(0)}\,(\mathrm{fm})$ & $1.348$ & $1.813$ & $1.802$ & $0.86$ & $2.792$ & $2.212$\tabularnewline
		\hline
	\end{tabular}

\end{table}

\begin{table}
	\centering
	\caption{Parameters of the potentials $U_{ij}^{(1)}$ for three different variants.}\label{tab:params1}
	\begin{tabular}{|l|c|c|c|c|}
		\hline
		& $u_{11}^{(1)}\,(\mathrm{MeV})$ & $u_{22}^{(1)}\,(\mathrm{MeV})$ & $u_{33}^{(1)}\,(\mathrm{MeV})$ & $a_{ii}^{(1)}\,(\mathrm{fm})$\tabularnewline
		\hline
		Variant I & $34.2$ & $-83.1$ & $0$ & $1.711$\tabularnewline
		Variant II & $58.3$ & $9.8$ & $-31.3$ & $1.584$\tabularnewline
		Variant III & $83.8$ & $167.1$ & $-39.6$ & $1.472$\tabularnewline
		\hline
	\end{tabular}
\end{table}

The parameters of the isoscalar potentials $U_{ij}^{(0)}$ that provide the best
description of the entire set of experimental data are given in Table~\ref{tab:params0}.
The values of these parameters are close to those obtained in our work~\cite{Salnikov2024}
without  account for the violation of isotopic invariance.
This is due to the fact that the cross sections of the processes $e^{+}e^{-}\to B\bar{B}$, $e^{+}e^{-}\to B^{*}\bar{B}$
and $e^{+}e^{-}\to B^{*}\bar{B}^{*}$, summed over charged
and neutral channels, weakly depend on the potentials $U_{ij}^{(1)}$
and are determined mainly by the potentials $U_{ij}^{(0)}$. Therefore, here
we  discuss only the effects of isotopic invariance violation,
that is, the deviation from unity of the ratios of the cross sections for pair production of charged and neutral mesons. For convenience, we introduce the notation $R_{mn}=\sigma^{(m)}/\sigma^{(n)}$,
\begin{equation}
R_{21}=\frac{\sigma(e^{+}e^{-}\to\Psi_{2})}{\sigma(e^{+}e^{-}\to\Psi_{1})}\,,\qquad R_{43}=\frac{\sigma(e^{+}e^{-}\to\Psi_{4})}{\sigma(e^{+}e^{-}\to\Psi_{3})}\,,\qquad R_{65}=\frac{\sigma(e^{+}e^{-}\to\Psi_{6})}{\sigma(e^{+}e^{-}\to\Psi_{5})}\,.\label{eq:R}
\end{equation}
The experimental data in Ref.~\cite{Abumusabh2025Measurements}
are obtained only for the  ratio $R_{21}$ in the vicinity of the resonance
$\Upsilon(4S)$. We verified that the influence of the potential $U_{33}^{(1)}$
and the off-diagonal potentials $U_{ij}^{(1)}$ on the ratio $R_{21}$
is small. However, taking into account the potential $U_{33}^{(1)}$ is important when analyzing
the ratios $R_{43}$ and $R_{65}$. Unfortunately, experimental information
for these ratios is extremely limited. Therefore, in order to demonstrate
the influence of isovector potentials on the ratios $R_{ij}$, we limited ourselves
to considering only the diagonal potentials $U_{11}^{(1)}$, $U_{22}^{(1)}$,
$U_{33}^{(1)}$ and set $U_{ij}^{(1)}=0$ for $i\ne j$.

\begin{figure}
\centering
\includegraphics[totalheight=5.7cm]{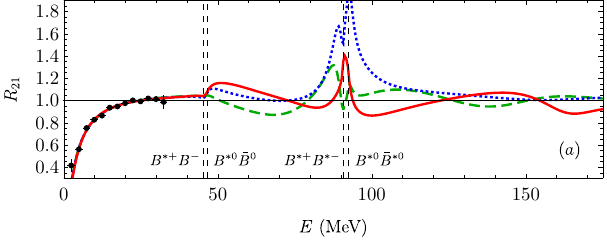}

\includegraphics[totalheight=5.7cm]{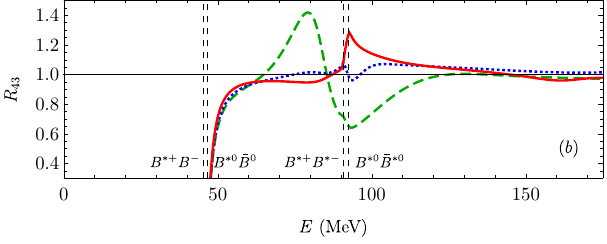}

\includegraphics[totalheight=5.7cm]{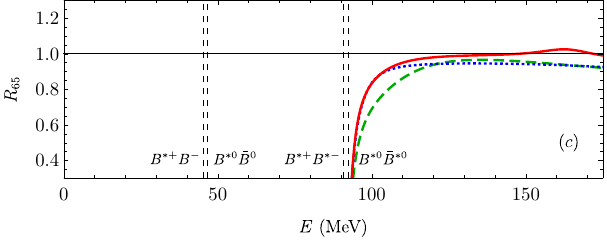}

\caption{Predictions for the energy dependence of  $R_{21}$, $R_{43}$
	and $R_{65}$. Solid lines correspond to variant I, dashed lines
	correspond to variant II, and dotted lines correspond to variant III
	from Table~\ref{tab:params1}. Experimental points for $R_{21}$
	are recalculated from data in Ref.~\cite{Abumusabh2025Measurements}.
	The vertical dashed lines indicate the thresholds for the production of the corresponding states.}\label{fig:1}
\end{figure}

Table~\ref{tab:params1} lists three sets of parameters for
$U_{ii}^{(1)}$. In all three cases, the model
describes well both the recent experimental data for $R_{21}$, obtained in Ref.~\cite{Abumusabh2025Measurements} in the vicinity of
$\Upsilon(4S)$, and the sums of the cross sections for the production of charged and
neutral particles obtained over the entire energy range under consideration~\cite{Aubert2009, Dong2020, Mizuk2021, Adachi2024Measurement}.
However, the predictions for $R_{21}$ at higher energies for different
parameter sets differ significantly and can reach large values (see Fig.~\ref{fig:1}(a)).
Large variations in $R_{21}$ are due to several factors. First, $R_{21}$ can differ significantly from unity near the $B^{*}\bar{B}$ and $B^{*}\bar{B}^{*}$ production thresholds.
Second, the $B\bar{B}$ production cross section above the $B^{*}\bar{B}$ threshold is significantly smaller than the $B^{*}\bar{B}$ and $B^{*}\bar{B}^{*}$ production cross sections. Therefore, due to nonzero transition amplitudes
between different channels, a small change due to isotopic invariance violation in the production amplitudes of $B^{*}\bar{B}$ and $B^{*}\bar{B}^{*}$
can lead to a significant change in the production amplitudes of $B^{+}B^{-}$
and $B^{0}\bar{B}^{0}$.
The second mechanism is most clearly manifested in the ratio  $R_{43}$ (see Fig.~\ref{fig:1}(b)). The large peak in $R_{43}$
in variant~II at an energy of about 75~MeV (15~MeV below the $B^{*}\bar{B}^{*}$ production threshold) is due to the fact that, according to our predictions~\cite{Salnikov2024},
for zero off-diagonal potentials a narrow bound
state exists in the $B^{*}\bar{B}^{*}$ channel at this energy. In the presence of transitions between
channels, this bound state decays into $B^{*}\bar{B}$ and
acquires a finite width. As a result, the value of $R_{43}$ becomes significantly
different from unity. This is despite the fact that the effects of isotopic invariance violation
(Coulomb interaction and mass differences between charged and neutral
$B^{(*)}$ mesons) are small compared to the scale of the strong
interaction potentials $U_{ij}^{(0)}$ and $U_{ij}^{(1)}$.
In variants I and III, this effect is not as pronounced, although even in these cases,
$R_{43}$ differs significantly from unity, Fig.~\ref{fig:1}(b). Although
the difference between  $R_{65}$  and unity is noticeably smaller than that for  $R_{21}$ and $R_{43}$, it can still reach
tens of percent, Fig.~\ref{fig:1}(c).

Note that there is  attempt in Ref.~\cite{Abumusabh2025Measurements}   to interpret the experimental data obtained for $R_{21}$  using a two-channel approximation for $B^{+}B^{-}$ and $B^{0}\bar B^{0}$ production. Our results confirm the conclusion of Ref.~\cite{Abumusabh2025Measurements} that the isovector potential in these channels is attractive.

\section{Conclusion}
We have analyzed the effects of isotopic invariance violation in the production of $B^{(*)}$ meson pairs in $e^{+}e^{-}$ annihilation in the energy range between  the $B^{+}B^{-}$ and $B_{s}^{0}\bar{B}_{s}^{0}$ production thresholds.
The analysis is based on the consideration of six-channel problem taking into account the final-state interaction in the $B^{(*)}\bar{B^{(*)}}$ system. Three channels correspond to the production of a pair of charged particles, and three channels correspond to the production of a pair of neutral particles. This approach yielded good agreement
with recent Belle~II~\cite{Abumusabh2025Measurements} results for the ratio of the $B^{0}\bar{B}^{0}$ and $B^{+}B^{-}$ production cross sections
in $e^{+}e^{-}$ annihilation in the vicinity of $\Upsilon(4S)$.

At higher energies, the predictions for each of the $R_{ij}$ ratios,
see Eq.~(\ref{eq:R}), can differ significantly from unity and from each other
for different sets of parameters. Therefore, detailed measurements of the effects
of isotopic invariance violation in $e^{+}e^{-}\to B^{(*)}\bar{B}^{(*)}$ processes in the energy range between $B^{+}B^{-}$ and $B_{s}^{0}\bar{B}_{s}^{0}$ production thresholds will  elucidate the structure of the interaction of
$B^{(*)}$ mesons. Observing a significant difference between  $R_{ij}$  and unity will serve as evidence that the complex energy dependence
 of the $B^{(*)}\bar{B}^{(*)}$ production cross sections in this energy range
is a consequence of the nontrivial interference of the particle production amplitudes
in the multichannel problem.

\subsection*{Acknowledgments}
We are grateful to A.E.~Bondar for useful discussions.


\begin{thebibliography}{99}
	\bibitem{Salnikov2024}
	S.G. Salnikov, A.E. Bondar, and A.I. Milstein, \href{https://dx.doi.org/10.1016/j.nuclphysa.2023.122764}{Nucl. Phys. A \textbf{1041}, 122764 (2024)}.
	\bibitem{Salnikov2024Production}
	S.G. Salnikov and A.I. Milstein, \href{https://dx.doi.org/10.1103/PhysRevD.109.114015}{Phys. Rev. D \textbf{109}, 114015 (2024)}.
	\bibitem{Aubert2004a}
	B. Aubert, et al. (BaBar Collaboration), \href{https://dx.doi.org/10.1103/PhysRevD.69.071101}{Phys. Rev. D \textbf{69}, 071101 (2004)}.
	\bibitem{Choudhury2023}
	S. Choudhury, et al. (Belle Collaboration), \href{https://dx.doi.org/10.1103/PhysRevD.107.L031102}{Phys. Rev. D \textbf{107}, L031102 (2023)}.
	\bibitem{Milstein2021}
	A.I. Milstein and S.G. Salnikov, \href{https://dx.doi.org/10.1103/PhysRevD.104.014007}{Phys. Rev. D \textbf{104}, 014007 (2021)}.
	\bibitem{Bondar2022}
	A.E. Bondar, A.I. Milstein, R.V. Mizuk, and S.G. Salnikov, \href{https://dx.doi.org/10.1007/JHEP05(2022)170}{J. High Energy Phys. \textbf{2022}, 170 (2022)}.
	\bibitem{Abumusabh2025Measurements}
	M. Abumusabh, et al., \href{https://arxiv.org/abs/2511.15926}{arXiv:2511.15926 [hep-ex]}.
	\bibitem{Navas2024Review}
	S. Navas, et al., \href{https://dx.doi.org/10.1103/PhysRevD.110.030001}{Phys. Rev. D \textbf{110}, 030001 (2024)}.
	\bibitem{Hayrapetyan2026Measurements}
	A.~Hayrapetyan, et al. (CMS Collaboration),
	\href{https://dx.doi.org/10.1103/njf9-4zfv}{Phys. Rev. Lett. \textbf{136} (2026)}.
	\bibitem{Salnikov2023}
	A.I. Milstein and S.G. Salnikov, \href{https://dx.doi.org/10.1134/S0021364023601471}{JETP Lett. \textbf{117}, 905 (2023)}.
	\bibitem{Mizuk2021}
	R. Mizuk, et al. (Belle Collaboration), \href{https://dx.doi.org/10.1007/JHEP06(2021)137}{J. High Energy Phys. \textbf{2021}, 137 (2021)}.
	\bibitem{Adachi2024Measurement}
	I. Adachi, et al. (Belle-II Collaboration), \href{https://dx.doi.org/10.1007/JHEP10(2024)114}{J. High Energ. Phys. \textbf{2024}, 114 (2024)}.
	\bibitem{Aubert2009}
	B. Aubert, et al. (BaBar Collaboration), \href{https://dx.doi.org/10.1103/PhysRevLett.102.012001}{Phys. Rev. Lett. \textbf{102}, 012001 (2009)}.
	\bibitem{Dong2020}
	X.-K. Dong, X.-H. Mo, P. Wang, and C.-Z. Yuan, \href{https://dx.doi.org/10.1088/1674-1137/44/8/083001}{Chin. Phys. C \textbf{44}, 083001 (2020)}.

\end{thebibliography}
\end{document}